# Off-axis MgB$_2$ films using an *in situ* annealing pulsed laser deposition method

Yue Zhao[1], Mihail Ionescu[2], Josip Horvat[1] and Shi Xue Dou[1]

[1] Institute for Superconducting and Electronic Materials, University of Wollongong, NSW 2522, Australia
[2] Australian Nuclear Science and Technology Organisation (ANSTO), NSW 2500, Australia
E-mail: yz70@uow.edu.au

**Abstract**: Highly smooth and *c*-axis oriented superconducting MgB$_2$ thin films were prepared by pulsed laser deposition (PLD) with off-axis geometry. The films were deposited on Al$_2$O$_3$-C substrates perpendicularly aligned to a stoichiometric MgB$_2$ target in a 120 mTorr high purity Ar background gas. An *in situ* annealing was carried out at 650 ℃ for 1 min in a 760 Torr Ar atmosphere. Despite the short annealing time, an x-ray θ-2θ scan shows fairly good crystallization, according to the clear *c*-axis oriented peaks for the films. Both atomic force microscopy and the x-ray diffraction results indicate that the crystallite size is less than 50nm. The root mean square roughness of our off-axis film is ~4 nm in a 5×5 μm$^2$ area. The $T_{c\,onset}$ value of the best off-axis film reaches 33.1 K with a narrow transition width of 0.9 K. The films showed no anisotropy in $H_{c2}$-$T$ curves when parallel and perpendicular fields were applied. The slope of $H_{c2}$-$T$ curves in low field regime is 1 T/K, which is among the highest reported values.

## 1. Introduction

In order to apply MgB$_2$ in superconducting electronics and coated conductors, various methods have been used for the preparation of MgB$_2$ films, such as physical vapor deposition (PVD) [1-5], chemical vapor deposition (CVD) [6], and electrochemical plating [7]. $T_c$'s of 39 K or even higher have been achieved in *ex situ* annealed PLD films [1] and Hybrid Physical-CVD (HPCVD) films [6]. The *in situ* annealed PLD films generally showed a considerably suppressed $T_c$ of ~25 K due to oxygen contamination and incomplete crystallization [8-11]. Although the $T_c$ value still requires improvement, the PLD procedure with an *in situ* annealing seems very attractive due to the fact that it requires comparatively simple equipment and is technically capable of preparing multi-layer films. Recent work on the *in situ* annealed PLD MgB$_2$ films in our group has shown that the $T_{c\,onset}$ can be significantly improved to 34.5 K, and the films show a combination of good superconducting properties [12].

However, the *in situ* annealed MgB$_2$ films prepared by PLD generally have many sub-micron and micron-sized particulates on their surfaces [10-12]. The bad surface quality of this type of film hinders its application in superconducting junctions or any other electronics. The presence of particulates on PLD films results from the splashing effect, which brings the melted droplets or detached fragments of the target to the films surface. Several methods have been developed to prevent particulates on PLD thin films [13, 14]. The probably most accepted solution is to apply an off-axis geometry, which places the substrate perpendicular to the target [15, 16]. The off-axis method has been proved effective and is easily adapted to PLD optical and electronic thin films, including HTS thin films [14-18]. However, we have not seen any report on off-axis deposition of superconducting MgB$_2$ films. In this paper we report on the growth of smooth MgB$_2$ thin films by PLD using off-axis geometry.

## 2. Experimental details

The PLD process was conducted in a spherical chamber with a volume of ~52 L. The stoichiometric MgB$_2$ target (84% density) and a magnesium target were set on a carousel in the chamber. Sapphire-C cut substrates with dimensions of about 6×2 mm$^2$ were used. For normal on-axis deposition, the substrate is mounted onto the resistive heater facing the target. In off-axis geometry, the substrate is parallel to the normal axis of the target surface and aligned to the center of the laser spot. It is mounted onto the edge of the heater with silver paste. The distance between the target and the substrate was about 25 mm for the off-axis deposition. A piece of metal screen was also attached to the heater just in front of the substrate to construct a shaded off-axis geometry [15], as shown in Figure 1.

The laser beam was generated by an excimer laser system (Lambda-Physik) operating on KrF gas ($\lambda$=248 nm, 25 ns). The laser beam was focused to an elliptical spot of ~7×1.5 mm$^2$ on the stoichiometric MgB$_2$ target with a density of 84%. The chamber was first evacuated to a base vacuum of ~8×10$^{-8}$ Torr and then filled with high purity argon to 120 mTorr as the background gas. During the deposition, the heater was kept at 250℃. At the end of the deposition, the Mg target was switched to the depositing position to provide a protective Mg cap-layer. This Mg layer was controlled to be ~800nm thick. Then the Ar pressure was increased to 760 Torr before the *in situ* annealing. The films were heated to 630-680℃ in 12 min and kept at that temperature for 1min. The power supply to the heater was then switched off, and the sample was cooled together with the heater at a cooling rate of approximately 50℃/min. The temperature difference between the film position and the control thermocouple was calibrated. An optimization of the annealing temperature was carried out, and the results show an optimal annealing temperature of about 650℃ for the off-axis films. The annealed off-axis films have a thickness of 400-600nm, as detected by atomic force microscopy (AFM, Digital Instruments)

The surface of the *in situ* annealed films was examined by scanning electron microscopy (SEM, Leica Cambridge), and the surface topography was further studied by AFM. We measured the temperature dependence of the resistivity in fields on a PPMS-9T magnetometer (Quantum Design) using a standard four-probe method and a dc current density of 10 A/cm$^2$. In both H//*ab*-plane and H⊥*ab*-plane cases, the testing current was perpendicular to the applied field. The hysteresis loops were measured on an MPMS-5T magnetometer with the applied field perpendicular to the film surface.

### 3. Results and discussions

We compare the surface of the off-axis films under different deposition conditions and an on-axis film in Fig. 2. For the on-axis deposited MgB$_2$ film, the film surface is covered with densely packed particulates, generally about one-micron in diameter. This kind of rough surface is quite common in the reported *in situ* annealed MgB$_2$ films [10, 11]. The off-axis deposited films generally do not contain many micron-sized particulates. However the surface roughness is not always low in off-axis deposited MgB$_2$ films. As shown in Fig. 2(b), under a comparatively high laser energy flux (500 mJ/pulse) and laser repetition frequency (10 Hz), the film surface looks loose and rough although few big particulates were found. At a lower laser energy flux (300 mJ/pulse) or smaller frequency (5 Hz), the films (films c and d respectively) are smoother. The long bumps observed on the surface of film (c) in Fig. 2 are droplets splashed from the MgB$_2$ target that landed along the film surface due to slight exposure of the substrate surface to the laser spot. With a shading screen, the long bumps were effectively prevented, as shown in Fig. 1(d). The number of micron-sized particulates on film (d) determined from SEM surface images is less than 1/100μm$^2$, which meets the general surface requirement for electronic films. In the remaining part of this paper, we only refer to the films prepared under the same conditions as film (d) as our off-axis films.

The smooth surface of the off-axis film is revealed in more detail by AFM, as shown in Fig. 3. The surface of the film is homogeneously constructed of round islands with diameters of about 60 nm. The nano-crystalline structure accompanied by a considerable volume of grain boundaries provides enhanced pinning to flux vortices [19], therefore the $J_c$ properties of the MgB$_2$ film might be improved. The average height of these round islands is 5nm resulting in a root mean square roughness ($R_q$) of 4 nm in a 5×5 μm area. The roughness of the film is comparable with the best MgB$_2$ films prepared by MBE [20] and HPCVD [6], which generally have $R_q$ values of 2-4nm.

The x-ray θ-2θ pattern of an off-axis film in Fig. 4 shows two peaks from MgB$_2$ (001) and (002), and a faint peak of Mg, indicating that the grains in the film are *c*–axis oriented and that the film contains a trace amount of free magnesium. The very quick annealing process of only one minute at 650℃ has triggered a reasonably good crystallization in the slowly deposited off-axis MgB$_2$ film. The full width at half maximum (FWHM) of MgB$_2$ (002) peak is broadened from 0.33° (the value of the target) to 0.68°. The broadening of the peaks is probably due to the small-grain feature and high level of disorder, which is also indicated by AFM and implied by the high residual resistivity of the film. The grain size calculated from the broadened FWHM using the Scherrer equation, without considering the inner strain, is about 25nm.

Figure 5 shows the temperature dependence of the resistivity for our best off-axis film. The $T_c$ onset of the film is 33.1 K, and the zero resistivity $T_c$ is 32.2 K, ~6 K below the values for bulk samples. This $T_c$ is close to that of the high quality *in situ* MBE MgB$_2$ films [5, 20], and among the highest reported values for *in situ* PLD films [8-12]. The suppression of $T_c$ for the films could be attributed to the lattice distortion resulted from the nano-grain structure or strong impurity scattering, especially inter-band scattering, which is pair-breaking in the two-gap superconductor. The resistivity difference $\Delta\rho_{300-40K}$ of the off-axis films is 8-9 μΩcm, comparatively low among MgB$_2$ films. Rowell *et al.* has argued that the resistivity difference between 300K and just above $T_c$ is indicative of the effective current carrying area [21]. The low $\Delta\rho_{300-40K}$ value probably shows a high density and good connection between the MgB$_2$ grains in the off-axis film. Thus the high residual resistivity of $\rho_{40K}$ = 30-60 μΩcm for the off-axis films is probably attributable to a high level of intra-grain disorders, which may also have considerably suppressed the $T_c$.

The inset figure of Fig. 5 shows the field dependence of the ρ-$T$ curves. We find that the superconducting transition width shows no significant broadening with increasing fields up to 8.7 T, which is quite different from what is observed in MgB$_2$ bulk samples [22], wires [23] and films [5], Since the resistivity transition width represents the regime of flux-flow resistivity [24], this result may indicate that the flux pinning in our *in situ* MgB$_2$ films remains consistent in both low and high fields.

The $H_{c2}$ curves, derived from transport curves using 90% $\rho_{Tc}$ values, reveal a high $H_{c2}$-$T$ slope, as shown in Fig. 6. The value of $dH_{c2}/dT$ near $T_c$ is about 1.0 T/K, for both the H//*ab*-plane and H⊥*ab*-plane circumstances. The fact that the $H_{c2}$ anisotropy of our *c*-oriented MgB$_2$ film is highly suppressed might be a result of a strong scattering, which may also dramatically increases the $H_{c2}$ value in the films. For dirty MgB$_2$ samples, there is no significant $H_{c2}$ saturation at low temperatures and the experimental $H_{c2}$-$T$ curves are approximately linear [25, 26]. Based on this, we roughly extrapolate that $H_{c2}(0) \approx$ 32 T at 0 K, using a linear equation, $H_{c2}(0)=H'_{c2}T_c$, where $H'_{c2}$ =|$dH_{c2}/dT$| near $T_c$. An estimation of the electron mean-free path $l$ using the equations for superconductors in GL theory, $H_{c2}(T) = \Phi_0/(2\pi\xi^2)$ and $\xi(T) = 0.855(\xi_0 l)^{1/2}/[1-T/T_c]^{1/2}$ (in the dirty limit), where $\Phi_0$ is the flux quantum and $\xi_0$ taken as the single crystal coherence length of ~6.0 nm [27], gives $l \approx$ 2.3 nm and $\xi \approx$ 3.1 nm at zero temperature.

The hysteresis measurements of the off-axis films reveal quite high $J_c$ values as well as a weak field dependence in low temperatures. Fig. 7 shows the calculated magnetic $J_c$ values of an off-axis film using the Bean model in perpendicular fields up to 5 T at different temperatures. The $J_c$ values in zero field are about 6×10$^6$ A/cm$^2$ at 10K, 3.5×10$^6$ A/cm$^2$ at 15K and 2×10$^6$ A/cm$^2$ at 20K. In a high field of 5 T, the $J_c$ value remains ~5×10$^5$ A/cm$^2$ at 5 K and ~2.3×10$^5$ A/cm$^2$ at 10 K. The good $J_c$ performance of the film is in agreement with the argument that the nano-grain structure may provide strong pinning. Also the above-mentioned possible high level of intra-grain disorders in the MgB$_2$ films may lead to a further pinning enhancement.

## 4. Conclusions

The off-axis deposition method is effective in improving the surface smoothness of PLD MgB$_2$ films. Under the selected deposition and annealing parameters, we obtained *c*-oriented *in situ* MgB$_2$ films on Al$_2$O$_3$-C substrate with a root mean square roughness of 4 nm. The $T_c$ onset of the best off-axis MgB$_2$ film is 33.1 K with a narrow transition width of 0.9 K. The combined high qualities of our off-axis films, namely good $T_c$ value, improved $J_c$ performance in high fields, high $H_{c2}$-$T$ slope of ~1 T/K and high surface quality, indicate that the off-axis PLD technique could be a viable candidate for the preparation of MgB$_2$ based electronics.

**Acknowledgements**

The authors thank T. Silver for many useful discussions. This work is supported by the Australian Research Council (ARC) under a Linkage Project (LP0219629) in cooperation with Alphatech International and The Hyper Tech Research Inc.

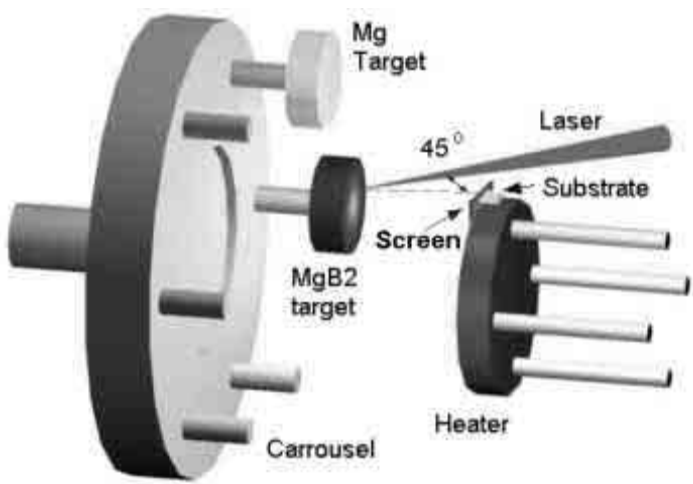

FIG. 1. Illustration of the off-axis deposition geometry for MgB$_2$ films

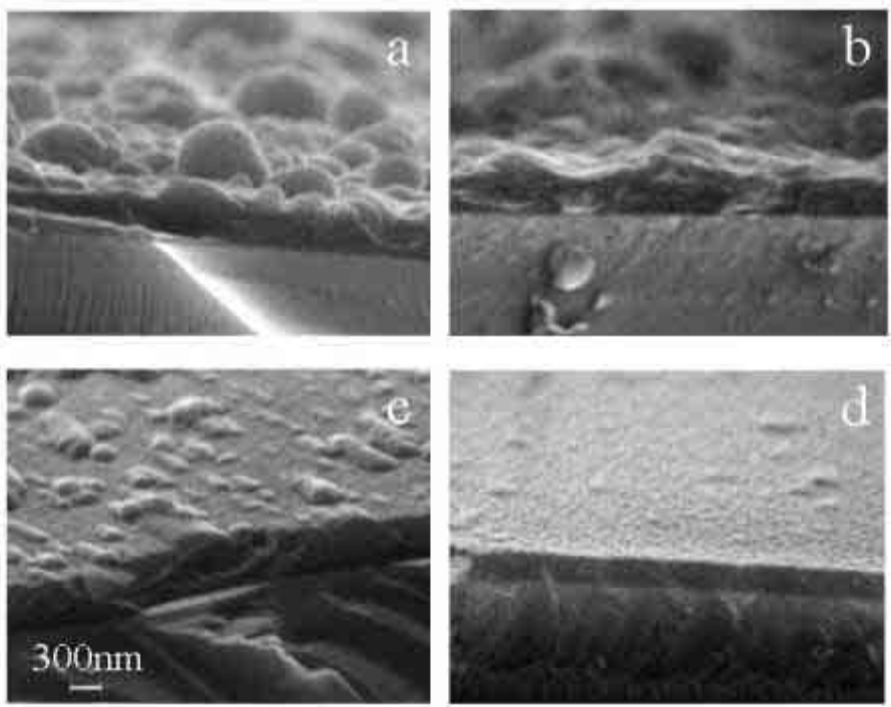

FIG. 2. SEM cross-sectional images of four films on $Al_2O_3$ substrates using different deposition conditions, namely (a): on-axis deposition, laser energy flux = 300 mJ/pulse, laser repetition frequency = 10 Hz, growth rate=12 Å/sec; (b) off-axis, E=500mJ/pulse, F=10Hz, 16Å/sec; (c) off-axis, E=300mJ/pulse, F=10Hz, 4 Å/sec; and (d) shaded off-axis, E=500mJ/pulse, F=5Hz, 2 Å/sec. The scale is the same for the four images.

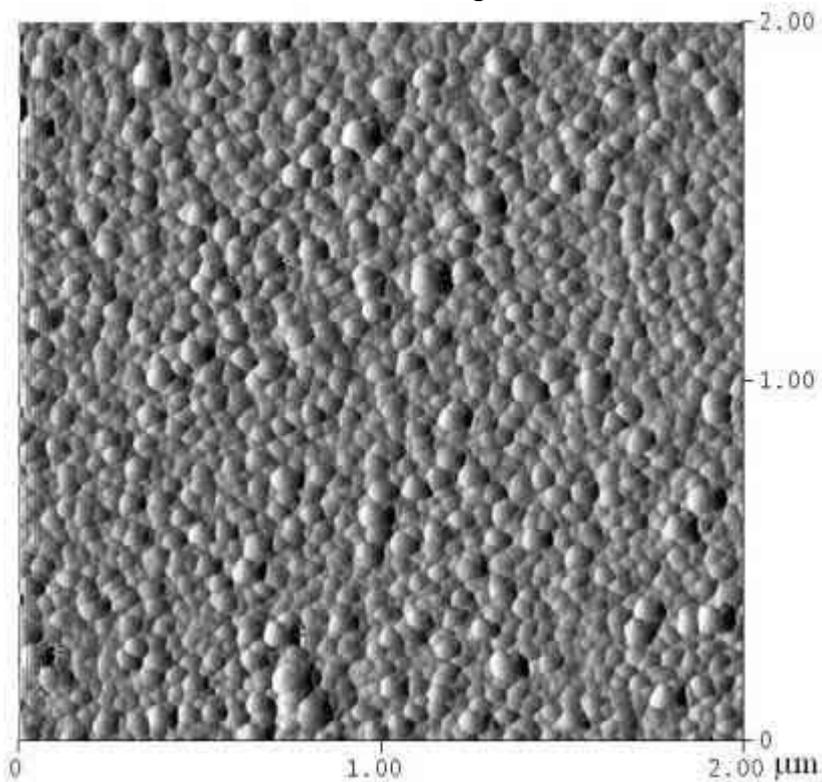

FIG. 3. AFM deflection image of a 2×2 $\mu m^2$ area of the surface of film (d) in Fig. 2.

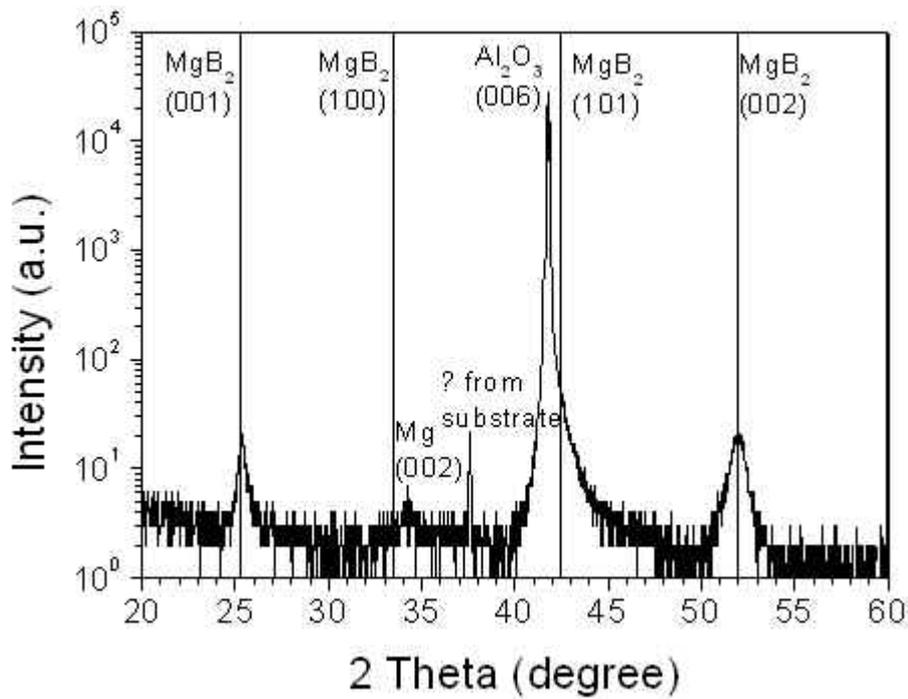

FIG. 4. XRD θ-2θ pattern of the off-axis deposited film with a slow scanning rate of 0.2 degree/min. The vertical lines label the positions for all $MgB_2$ peaks in powder diffraction database. The unknown peak at 37.56º also presents in the spectrum for bare $Al_2O_3$-C substrate, so we assume it is not from the film.

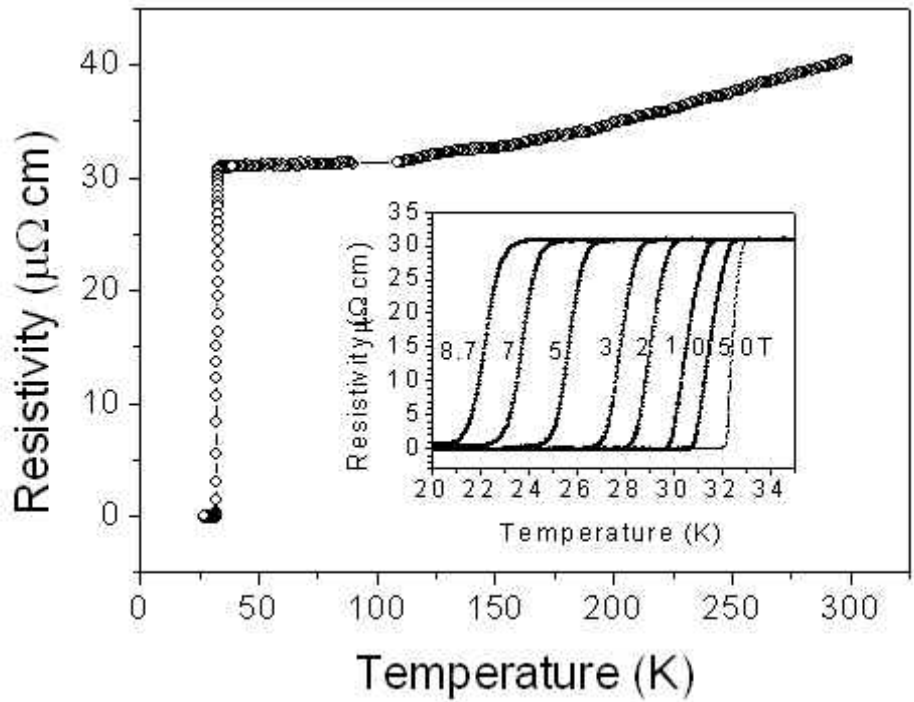

FIG. 5. Resistivity versus temperature for an off-axis film. Inset figure shows the field dependence of ρ-T curves with H ⊥ ab-plane.

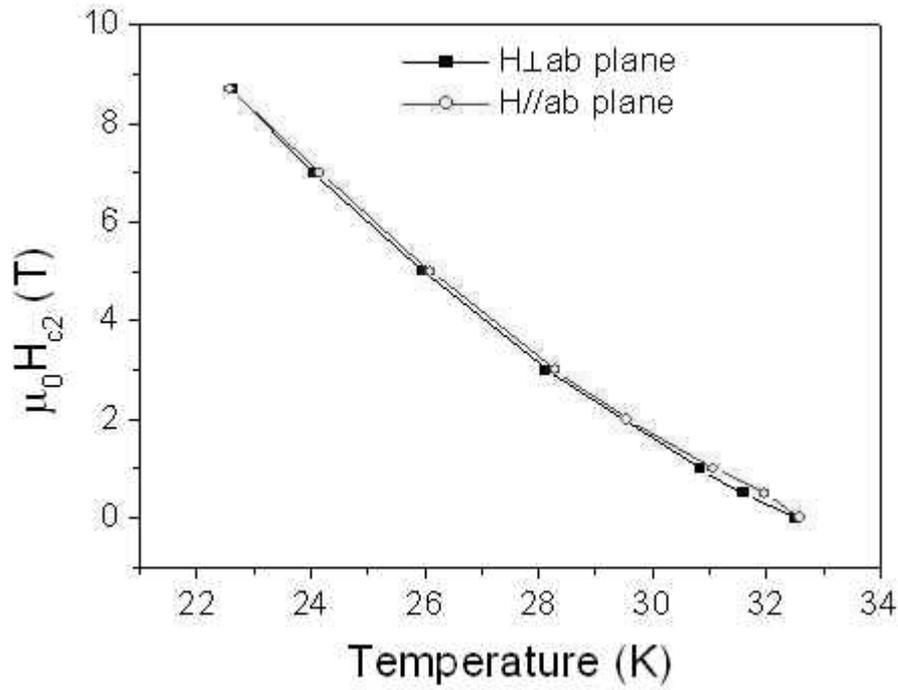

FIG. 6. The $H_{c2}$-T curves for H // $ab$-plane and H $\perp$ $ab$-plane. The $H_{c2}$ values are derived from transport curves using 90% $\rho_{Tc}$ values. In both H // $ab$-plane and H$\perp$$ab$-plane cases, the testing current was perpendicular to the applied field.

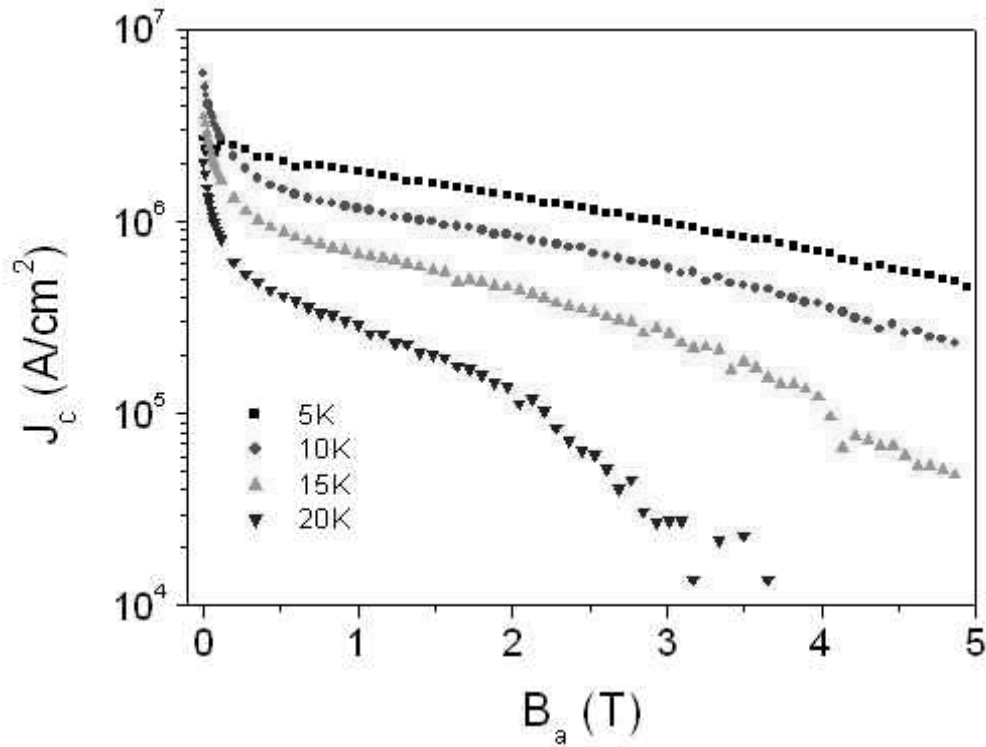

FIG. 7. Magnetic $J_c$ versus applied field for an off-axis film at different temperatures. It is difficult to estimate $J_c$ at 5 T in low fields owing to the predominant magneto-thermal instability.